\def\year{2022}\relax
\documentclass[letterpaper]{article} %
\usepackage{aaai22}  %
\usepackage{times}  %
\usepackage{helvet}  %
\usepackage{courier}  %
\usepackage[hyphens]{url}  %
\usepackage{graphicx} %
\usepackage{natbib}  %
\usepackage{caption} %
\DeclareCaptionStyle{ruled}{labelfont=normalfont,labelsep=colon,strut=off} %
\usepackage{floatrow}
\usepackage{times}
\usepackage{epsfig}
\usepackage{amsmath}
\usepackage{amssymb}
\usepackage{algorithm}
\usepackage{algorithmic}
\usepackage{subcaption}
\usepackage{booktabs}
\usepackage{multirow}
\usepackage{subcaption}
\newcommand{\eg}{\textit{e}.\textit{g}.}

\usepackage{newfloat}
\usepackage{listings}
\floatstyle{ruled}
\newfloat{listing}{tb}{lst}{}
\floatname{listing}{Listing}

\title{Parallel and High-Fidelity Text-to-Lip Generation}
\author{
    Jinglin Liu\thanks{Equal contribution.}\textsuperscript{\rm 1},
    Zhiying Zhu\footnotemark[1]\textsuperscript{\rm 1},
    Yi Ren\footnotemark[1]\textsuperscript{\rm 1}, \\
    Wencan Huang\textsuperscript{\rm 1},
    Baoxing Huai\textsuperscript{\rm 2},
    Nicholas Yuan\textsuperscript{\rm 2},
    Zhou Zhao\thanks{Corresponding author}\textsuperscript{\rm 1}
}
\affiliations{
    \textsuperscript{\rm 1}Zhejiang University, China \\
    \textsuperscript{\rm 2}Huawei Cloud \\
    \{jinglinliu,zhyingzh,rayeren,huangwencan,zhaozhou\}@zju.edu.cn,\\
    \{huaibaoxing,nicholas.yuan\}@huawei.com

}

\usepackage{bibentry}

\begin{document}
\maketitle
\begin{abstract}
   As a key component of talking face generation, lip movements generation determines the naturalness and coherence of the generated talking face video. Prior literature mainly focuses on speech-to-lip generation while there is a paucity in text-to-lip (T2L) generation. T2L is a challenging task and existing end-to-end works depend on the attention mechanism and autoregressive (AR) decoding manner. However, the AR decoding manner generates current lip frame conditioned on frames generated previously, which inherently hinders the inference speed, and also has a detrimental effect on the quality of generated lip frames due to error propagation. This encourages the research of parallel T2L generation. In this work, we propose a parallel decoding model for fast and high-fidelity text-to-lip generation (ParaLip). Specifically, we predict the duration of the encoded linguistic features and model the target lip frames conditioned on the encoded linguistic features with their duration in a non-autoregressive manner. Furthermore, we incorporate the structural similarity index loss and adversarial learning to improve perceptual quality of generated lip frames and alleviate the blurry prediction problem. Extensive experiments conducted on GRID and TCD-TIMIT datasets demonstrate the superiority of proposed methods. Video samples are available via \url{https://paralip.github.io/}.
   
\end{abstract}

\section{Introduction}
In the modern service industries, talking face generation has broad application prospects such as avatar, virtual assistant, movie animation, teleconferencing, etc. \cite{zhu2020Arbitrary}. As a key component of talking face generation, lip movements generation (a.k.a. lip generation) determines the naturalness and coherence of the generated talking face video. Lip generation aims to synthesize accurate mouth movements video corresponding to the linguistic content information carried in speech or pure text. 

\begin{figure}[!t]
	\centering
	
	\includegraphics[width=\textwidth]{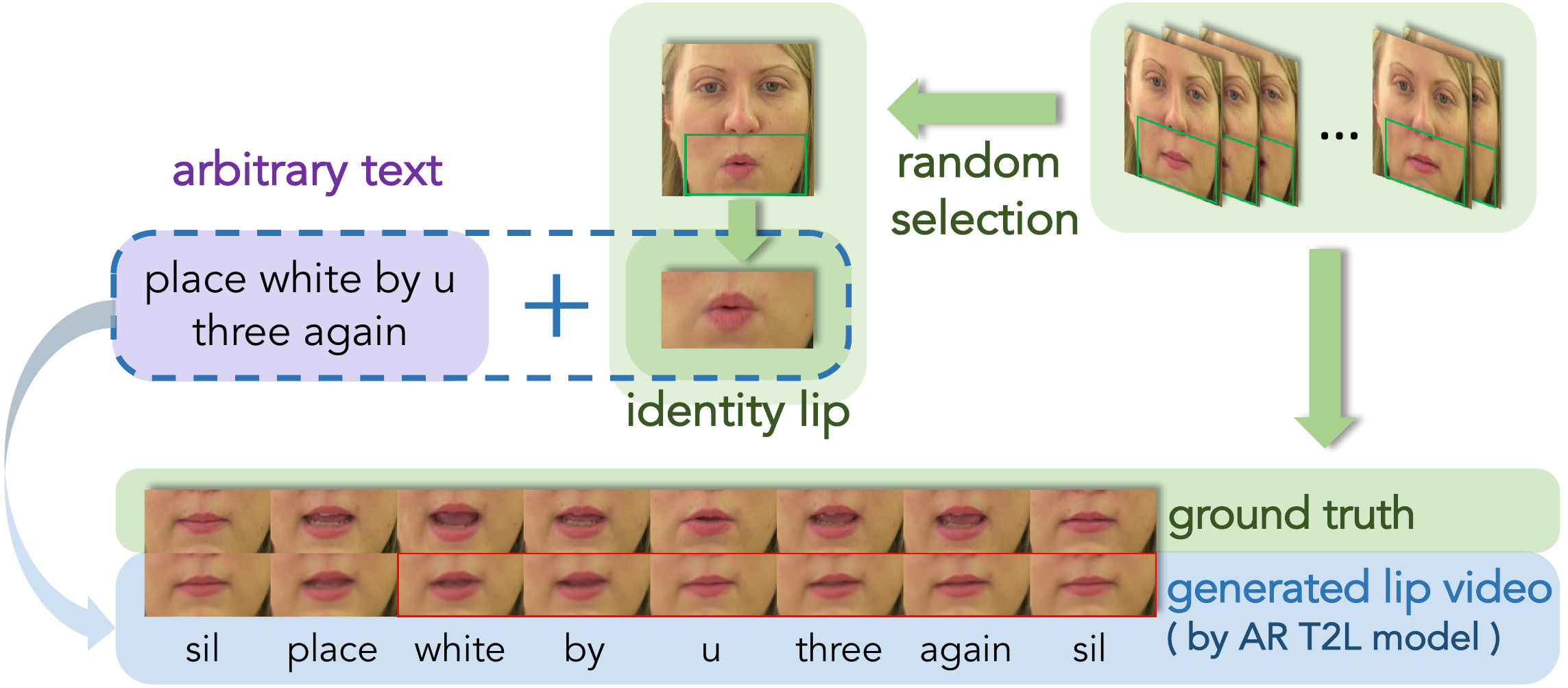}
	\small
	\caption{The task description of T2L generation. The model takes in an arbitrary source text sequence and a single identity lip image to synthesize the target lip movements video. And in this figure we can see that the generated video loses the linguistic information gradually and finally becomes fuzzy and motionless (lip frames in the red box), which is the intractable problem existing in AR T2L models due to error propagation.}
	\label{fig:intro_fig}
\end{figure}

Mainstream literature focuses on speech-to-lip (S2L) generation while there is a paucity in text-to-lip (T2L) generation. Even so, T2L generation is very crucial and has considerable merits compared to S2L since 1) text data can be obtained or edited more easily than speech, which makes T2L generation more convenient; and 2) T2L extremely preserves privacy especially in the society where the deep learning techniques are so developed that a single sentence speech could expose an unimaginable amount of personal information.

However, end-to-end T2L task (shown in Figure \ref{fig:intro_fig}) is challenging. Unlike S2L task where the mapping relationship between the sequence length of source speech and target video is certain (according to audio sample rate and fps), there is an uncertain sequence length discrepancy between source and target in T2L task. The traditional temporal convolutional networks become impractical. Hence, existing works view T2L as a sequence-to-sequence task and tackle it by leveraging the attention mechanism and autoregressive (AR) decoding manner. The AR decoding manner brings two drawbacks: 1) it inherently hinders the inference speed since its decoder generates target lips one by one autoregressively with the causal structure. Consequently, generating a single sentence of short video consumes about 0.5-1.5 seconds even on GPU, which is not acceptable for industrial applications such as real-time interactions with avatar or virtual assistant, real-time teleconferencing and document-level audio-visual speech synthesis, etc. 2) It has a detrimental effect on the quality of generated lips due to error propagation\footnote{Error propagation means if a token is mistakenly predicted at inference stage, the error will be propagated and the future tokens conditioned on this one will be influenced~\cite{bengio2015scheduled,wu-etal-2018-beyond}.}, which is frequently discussed in neural machine translation and image caption field~\cite{bengio2015scheduled,wu-etal-2018-beyond}. Worse still, error propagation is more obvious in AR lip generation than in other tasks, because the mistakes could take place at more dimensions (every pixel with three channels in generated image) and there is information loss during the down-sampling when sending the last generated lip frame to predict current one. Although prior works alleviate the error propagation by incorporating the technique of location-sensitive attention, it still has an unsatisfying performance on long-sequence datasets due to accumulated prediction error.

To address such limitations, we turn to non-autoregressive (NAR) approaches. NAR decoding manner generates all the target tokens in parallel, which has already pervaded multiple research fields such as neural machine translation~\cite{gu2018non,lee2018deterministic,ghazvininejad2019mask,ma2019flowseq}, speech recognition~\cite{chen2019non,higuchi2020mask}, speech synthesis~\cite{ren2019fastspeech,peng2020non,miao2020flow}, image captioning~\cite{deng2020length} and lip reading~\cite{liu2020fastlr}. These works utilize the NAR decoding in sequence-to-sequence tasks to reduce the inference latency or generate length-controllable sequence.

In this work, we propose an NAR model for parallel and high-fidelity T2L generation (ParaLip). ParaLip predicts the duration of the encoded linguistic features and models the target lip frames conditioned on the encoded linguistic features with their duration in a non-autoregressive manner. Furthermore, we leverage structural similarity index (SSIM) loss to supervise ParaLip generating lips with better perceptual quality. Finally, using only reconstruction loss and SSIM loss is insufficient to generate distinct lip images with more realistic texture and local details (\eg wrinkles, beard and teeth), and therefore we adopt adversarial learning to mitigate this problem.

Our main contributions can be summarized as follows: 1) We point out and analyze the unacceptable inference latency and intractable error propagation existing in AR T2L generation. 2) To circumvent these problems, we propose ParaLip to generate high-quality lips with low inference latency. And as a byproduct of ParaLip, the duration predictor in ParaLip could be leveraged in an NAR text-to-speech model, which naturally enables the synchronization in audio-visual speech synthesis task. 3) We explore the source-target alignment method when the audio is absent even in the training set. Extensive experiments demonstrate that ParaLip generates the competitive lip movements quality compared with state-of-the-art AR T2L model and exceeds the baseline AR model TransformerT2L by a notable margin. In the meanwhile, ParaLip exhibits distinct superiority in inference speed, which truly provides the possibility to bring T2L generation from laboratory to industrial applications.

\section{Related Work}
\subsection{Talking Face Generation}
Talking face generation aims to generate realistic talking face video and covers many applications such as avatar and movie animation. There is a branch of works in computer graphics (CG) field exploring it \cite{wang2011text,fan2015photo,suwajanakorn2017synthesizing,yu2019durian,abdelaziz2020audiovisual} through hidden Markov models or deep neural network. These works synthesize the whole face by generating the intermediate parameters, which can then be used to deform a 3D face. Thanks to the evolved convolutional neural network (CNN) and high-performance computing resources, end-to-end systems which synthesize 2D talking face images by CNN rather than rendering methods in CG, have been presented recently in the computer vision (CV) field \cite{kumar2017obamanet,chung2017you,Chen2018Lip,vougioukas2019end,zhou2019talking,zhu2020Arbitrary,zheng2020photorealistic,prajwal2020lip}. Most of them focus on synthesizing lip movements images and then transforming them to faces. We mainly take these works in CV field into consideration and broadly divide them into two streams as the following paragraphs. 

\paragraph{Speech-to-Lip Generation}
Previous speech-driven works \eg  \citet{chung2017you} simply generate the talking face images conditioned on the encoded speech and the encoded face image carrying the identity information. To synthesize more accurate and distinct lip movements, \citet{Chen2018Lip} introduce the task of speech-to-lip generation using lip image as the identity information. Further, \citet{song2018talking} add a lip-reading discriminator to focus on the mouth region, and \citet{zhu2020Arbitrary} add the dynamic attention on lip area to synthesize talking face while keeping the lip movements realistic. \citet{prajwal2020lip} propose a pre-trained lip-syncing discriminator to synthesize talking face with speech-consistent lip movements. 

\paragraph{Text-to-Lip Generation}
The literature of direct text-to-lip generation is rare. Some text-driven approaches either cascade the text-to-speech and speech-to-lip generation model\cite{kr2019towards,kumar2017obamanet}, or combine the text feature with speech feature together to synthesize lip movements \cite{yu2019mining}. \citet{fried2019text} edit a given video based on pure speech-aligned text sequence. Unlike the scenario where source speech or video is given, the sequence length of target lip frames is uncertain with only text input. Existing work \cite{chen2020DualLip} depends on the attention mechanism and AR decoding method to generate the target lip frames until the stop token is predicted.

\begin{figure*}[!ht]
  
  \begin{floatrow}
   
    \ffigbox[\textwidth]{
      \begin{subfloatrow}[4]
        \centering
        \ffigbox[0.873\FBwidth]{
        
          \includegraphics[height=5.9cm]{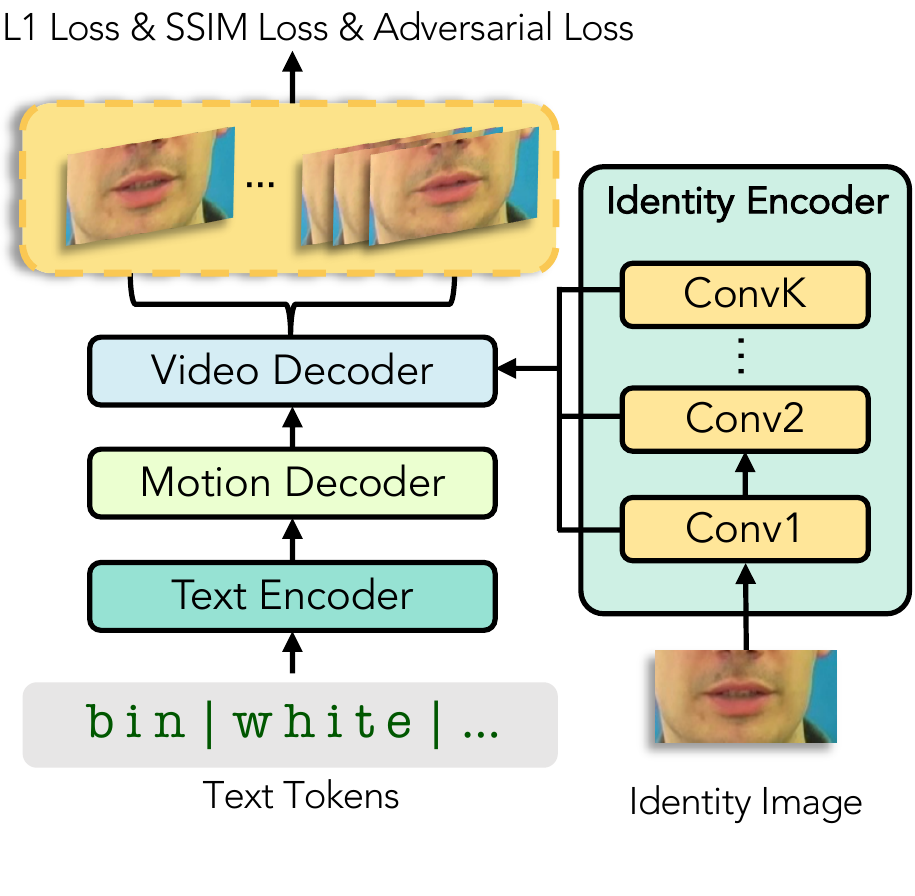}
        }{\caption{ParaLip.}\label{fig:hht2l_overview}}
        \centering
        \ffigbox[0.782\FBwidth]{
            
         	\includegraphics[height=5.9cm]{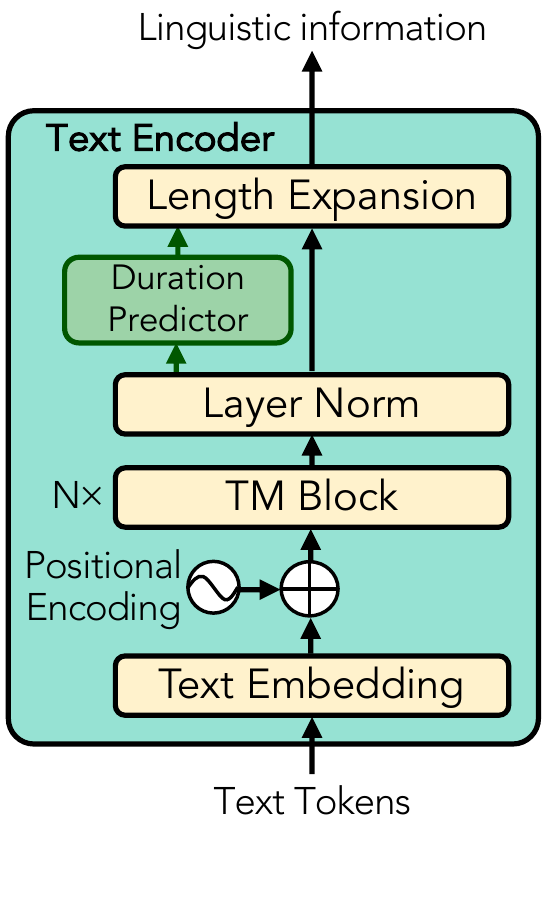}
        }{\caption{Text Encoder with Length Regulator.}\label{fig:text_encoder}}
        \centering
        \ffigbox[0.70\FBwidth]{
         	\includegraphics[height=5.9cm]{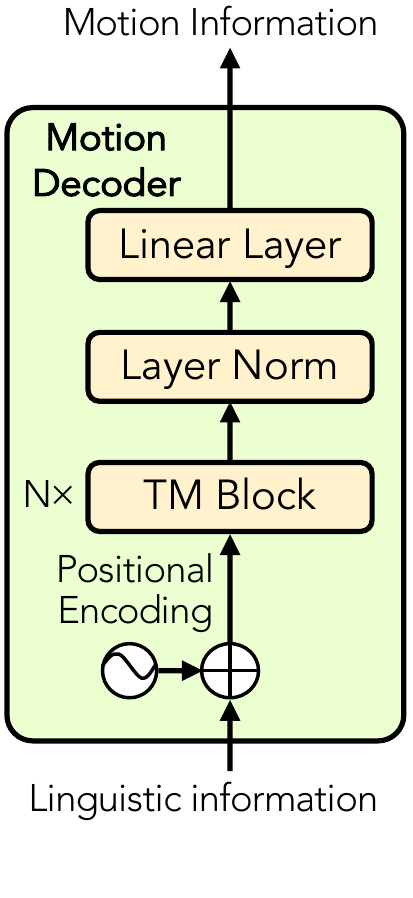}
        }{\caption{Motion Decoder.}\label{fig:motion_decoder}}
        \centering
        \ffigbox[0.81\FBwidth]{
         	\includegraphics[height=5.9cm]{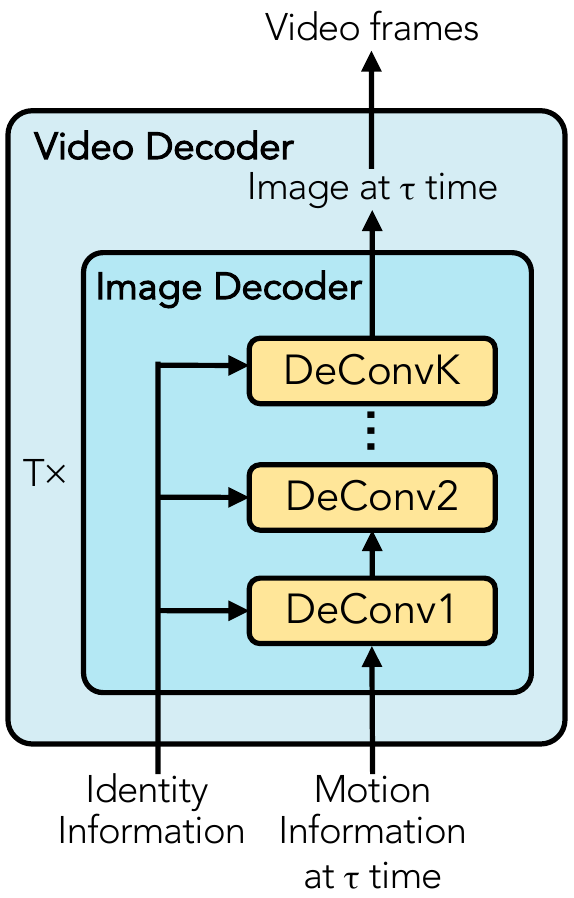}
        }{\caption{Video Decoder with multiple Image Decoders.} \label{fig:video_decoder}}
        \end{subfloatrow}
    }{\caption{The overall architecture for ParaLip. In subfigure (a), Identity Encoder sends out residual information at every convolutional layer. In subfigure (b), Length Regulator expands the text sequence according to ground truth duration in training or predicted duration in inference. In subfigure (c), Motion Decoder models lip movement information sequence from linguistic information sequence. In subfigure (d), there are $T$ Image Decoders placed parallel in Video Decoder. The $\tau$-th Image Decoder takes in motion information at $\tau$ time and generates lip image at $\tau$ time. $T$ means total number of lip frames.}\label{fig:ParaLip_design}}
  \end{floatrow}
\end{figure*}

\subsection{Non-Autoregressive Sequence Generation}
In sequence-to-sequence tasks, an autoregressive (AR) model takes in a source sequence and then generates tokens of the target sentence one by one with the causal structure at inference \cite{sutskever2014sequence,vaswani2017attention}. Since the AR decoding manner causes the high inference latency, many non-autoregressive (NAR) models, which generate target tokens conditionally independent of each other, have been proposed recently. Earliest in the NAR machine translation field, many works use the fertility module or length predictor \cite{gu2018non,lee2018deterministic,ghazvininejad2019mask,ma2019flowseq} to predict the length correspondence (fertility) between source and target sequences, and then generate the target sequence depending on the source sequence and predicted fertility. Shortly afterward, researchers bring NAR decoding manner into heterogeneous tasks. In the speech field, NAR-based TTS \cite{ren2019fastspeech,peng2020non,miao2020flow} synthesize speech from text with high speed and slightly quality drop; NAR-based ASR \cite{chen2019non,higuchi2020mask} recognize speech to corresponding transcription faster. In the computer vision field, \citet{liu2020fastlr} propose an NAR model for lipreading; \citet{deng2020length} present NAR image caption model not only improving the decoding efficiency but also making the generated captions more controllable and diverse.

\section{Method}

\subsection{Preliminary Knowledge}
The text-to-lip generation aims to generate the sequence of lip movement video frames $\mathcal{L} = \{l_1, l_2, ..., l_T\}$, given source text sequence $\mathcal{S} = \{s_1, s_2, ..., s_m\}$ and a single identity lip image $l_{I}$ as condition. Generally, there is a considerable discrepancy between the sequence length of $\mathcal{L}$ and $\mathcal{S}$ with uncertain mapping relationship. Previous work views this as a sequence-to-sequence problem, utilizing attention mechanism and AR decoding manner, where the conditional probability of $\mathcal{L}$ can be formulated as: 
\begin{equation}
    P(\mathcal{L}|\mathcal{S}, l_{I}) = \prod_{\tau=0}^{T}P(l_{\tau+1} | l_{< \tau+1}, \mathcal{S}, l_{I}; \theta), 
\end{equation}
where $\theta$ denotes the parameters of the model.

To remedy the error propagation and high latency problem brought by AR decoding, ParaLip models the target sequence in an NAR manner, where the conditional probability becomes:
\begin{equation}
    P(\mathcal{L}|\mathcal{S}, l_{I}) = \prod_{\tau=1}^{T} P(l_{\tau}| \mathcal{S}, l_{I}; \theta).
\end{equation}

\subsection{Model Architecture of ParaLip}
The overall model architecture and training losses are shown in Figure \ref{fig:hht2l_overview}. We explain each component in ParaLip in the following paragraphs. 

\paragraph{Identity Encoder}
As shown in the right panel of Figure \ref{fig:hht2l_overview}, identity encoder consists of stacked 2D convolutional layers with batch normalization, which down-samples the identity image multiple times to extract features. The identity image is selected randomly from target lip frames, providing the appearance information of a speaker. It is worth noting that the identity encoder sends out the final encoded hidden feature together with the intermediate hidden feature of convolutional layers at every level, which provides the fine-grained image information. 

\paragraph{Text Encoder}
As shown in  Figure \ref{fig:text_encoder}, the text encoder consists of a text embedding layer, stacked feed-forward Transformer layers (TM) \cite{ren2019fastspeech}, a duration predictor and a length regulator. The TM layer contains self-attention layer and 1D convolutional layer with layer normalization and residual connection \cite{vaswani2017attention,gehring2017convolutional}. The duration predictor contains two 1D convolutional layers with layer normalization and one linear layer, which takes in the hidden text embedding sequence and predicts duration sequence $\mathcal{D}^* = \{d^*_1, d^*_2, ..., d^*_m\}$, where $d^*_i$ means how many video frames the $i$-th text token corresponding to. The length regulator expands the hidden text embedding sequence according to ground truth duration $\mathcal{D}$ at training stage or predicted duration $\mathcal{D}^*$ at inference stage. For example, when given source text and duration sequence are $\{s_1, s_2, s_3\}$ and $\{2,1,3\}$ respectively, denoting the hidden text embedding as $\{h_1, h_2, h_3 \}$,  the expanded sequence is $\{h_1, h_1, h_2, h_3, h_3, h_3\}$, which carries the linguistic information corresponding to lip movement video at frame level. Collectively, text encoder encodes the source text sequence $S$ to linguistic information sequence $\widetilde{\mathcal{S}} = \{\widetilde{s}_1, \widetilde{s}_2, ..., \widetilde{s}_{T^*} \}$, where $T^* = \sum_{i=1}^{m}d_i$ at training stage, or $T^* = \sum_{i=1}^{m}d^*_i$ at inference stage.

\paragraph{Motion Decoder}
Motion decoder (Figure \ref{fig:motion_decoder}) aims to model the lip movement information sequence $\widetilde{\mathcal{L}}=\{\widetilde{l}_1, \widetilde{l}_2, ..., \widetilde{l}_{T^*}\} $ from linguistic information sequence $\widetilde{\mathcal{S}}$. It utilizes the positional encoding and self-attention mechanism in stacked TM blocks to enforce the temporal correlation on the hidden sequence. There is a linear layer at the end of this module convert the hidden states to an appropriate dimension. %

\paragraph{Video Decoder}
The video decoder generates the target lip movement video $\mathcal{L}^*$ conditioned on the motion information sequence and identity information. As shown in Figure \ref{fig:video_decoder}, the video decoder consists of multiple parallel image decoders with all parameters shared, each of which contains stacked 2D deconvolutional layers, and there are skip connections at every level between the identity encoder and each image decoder. The skip connection is implemented by concatenation. Then two extra 2D convolutional layers are added at the end of each decoder for spatial coherence. Finally, the $\tau$-th image decoder takes in lip motion information at $\tau$ time $\widetilde{l}_{\tau}$ and generates lip image $l^*_{\tau}$ at $\tau$ time in corresponding shape. 

\subsection{Training Methods}
In this section, we describe the loss function and training strategy to supervise ParaLip. The reconstruction loss and duration prediction loss endow the model with the fundamental ability to generate lip movement video. To generate the lip with better perceptual quality and alleviate the ``blurry predictions''~\cite{mathieu2016deep} problem, the structural similarity index loss and adversarial learning are introduced. We also explore the source-target alignment method when the audio is absent even in the training set, which will be introduced in Section~\ref{sec:further_discussion}.

\paragraph{Reconstruction Loss}
Basically, we optimize the whole network by adding $L_1$ reconstruction loss on generated lip sequence $\mathcal{L}^*$: 
\begin{equation}
    L_{rec} = \sum_{\tau=1}^T \| l_{\tau} - l^*_{\tau} \| _{1}.
\end{equation}

\paragraph{Duration Prediction Loss}
In the training stage, we add $L_1$ loss on predicted duration sequence $\mathcal{D}^*$ at token level\footnote{Character level for GRID and phoneme level for TCD-TIMIT following previous works.} and sequence level, which supervises the duration predictor to make the precise fine-grained and coarse-grained predictions. Duration prediction loss $L_{dur}$ can be written as:
\begin{equation}
    L_{dur} = \sum_{i=1}^m \| d_{i} - d^*_i \| _{1} + \| \sum_{i=1}^m d_i - \sum_{i=1}^m d_i^* \| _{1}.
\end{equation}

\paragraph{Structural Similarity Index Loss}
Structural Similarity Index (SSIM)~\cite{Wang2004Image} is adopted to measure the perceptual image quality, which takes luminance, contrast and structure into account, and is close to the perception of human beings. The SSIM value for two pixels at position $(i, j)$ in $\tau$-th images $l^*_{\tau}$ and $l_{\tau}$ can be formulated as:
\begin{equation*}
SSIM_{i, j, \tau}=\frac{2 \mu_{l^*_{\tau}} \mu_{l_{\tau}}+C_{1}}{\mu_{l^*_{\tau}}^{2}+\mu_{l_{\tau}}^{2}+C_{1}} \cdot \frac{2 \sigma_{l^*_{\tau} l_{\tau}}+C_{2}}{\sigma_{l^*_{\tau}}^{2}+\sigma_{l_{\tau}}^{2}+C_{2}},
\end{equation*}
where $\mu_{l^*_{\tau}}$ and $\mu_{l_{\tau}}$ denotes the mean for regions in image $l^*_{\tau}$ and $l_{\tau}$ within a 2D-window surrounding $(i, j)$. Similar, $\sigma_{l^*_{\tau}}$ and $\sigma_{l_{\tau}}$ are standard deviation; $\sigma_{l^*_{\tau} l_{\tau}}$ is the covariance; $C_{1}$ and $C_{2}$ are constant values. To improve the perceptual quality of the generated lip frames, we leverage SSIM loss in ParaLip. Assuming the size of each lip frame to be $(A \times B)$, the SSIM loss between generated $\mathcal{L}^*$ and ground truth $\mathcal{L}$ becomes:
\begin{equation}
    L_{ssim}=\frac{1}{T\cdot A\cdot B}\sum_{\tau=1}^T\sum_{i}^{A}\sum_{j}^{B}(1-SSIM_{i,j,\tau})).
\end{equation}

\paragraph{Adversarial Learning}
Through experiments, it can be found that only using above losses is insufficient to generate distinct lip images with more realistic texture and local details (\eg wrinkles, beard and teeth). Thus, we adopt adversarial learning to mitigate this problem and train a quality discriminator $Disc$ along with ParaLip. The $Disc$ contains stacked 2D convolutional layers with LeakyReLU activation which down-samples each image to $1\times1\times H$ ($H$ is hidden size), and a $1\times1$ convolutional layer to project the hidden states to a value of probability for judging real or fake. We use the loss function in LSGAN \cite{mao2017least} to train ParaLip and $Disc$:
\begin{equation}
    L_{adv}^G = \mathbb{E}_{x \sim l^*}(Disc(x)-1)^{2},
\end{equation}
\begin{equation}
\label{eq:adv_d}
    L_{adv}^D = \mathbb{E}_{x \sim l}(Disc(x)-1)^{2}+\mathbb{E}_{x \sim l^*}Disc(x)^{2},
\end{equation}
where $l^*$ means lip images generated by ParaLip and $l$ means ground truth lip images. 

To summarize, we optimize the $Disc$ by minimizing Equation \eqref{eq:adv_d}, and optimize the ParaLip by minimizing $L_{total}$:
\begin{equation}
    L_{total} = \lambda_1 \cdot L_{rec} + \lambda_2 \cdot L_{dur} + \lambda_3 \cdot L_{ssim} + \lambda_4 \cdot L_{adv}^G,
\end{equation}
where the $\lambda_1$, $\lambda_2$, $\lambda_3$ and $\lambda_4$ are hyperparameters to trade off
the four losses.

\section{Experimental Settings}

\subsection{Datasets}
\paragraph{GRID}
The GRID dataset~\cite{griddataset} consists of 33 video-available speakers, and each speaker utters 1,000 phrases. The phrases are in a 6-categories structure following fixed simple grammar: $command^{4} + color^{4} + preposition^{4} + letter^{25} + digit^{10} + adverb^{4}$ where the number denotes how many choices of each category. Thus, the total vocabulary size is 51, composing 64,000 possible phrases. All the videos last 3 seconds with frame rate 25 fps, which form a total duration of 27.5 hours. It is a typical talking face dataset and there are a considerable of lip-related works \cite{Assael2016LipNet,chung2017lip,afouras2018deep,Chen2018Lip,zhu2020Arbitrary,10.1145/3474085.3475220} conducting experiments on it. Following previous works, we select 255 random samples from each speaker to form the test set.

\paragraph{TCD-TIMIT}
The TCD-TIMIT dataset~\cite{Harte2015TCD} is closer to real cases and more challenging than GRID dataset, since 1) the vocabulary is not limited; 2) the sequence length of videos is not fixed and is longer than that in GRID. We use the `volunteers' subset of TCD-TIMIT following previous works, which consists of 59 speakers uttering about 98 sentences individually. The frame rate is 29.97 fps and each video lasts 2.5$ \sim $8.1 seconds. The total duration is about 7.5 hours. 
We set 30\% of data from each speaker aside for testing following the recommended speaker-dependent train-test splits~\cite{Harte2015TCD}.

\subsection{Data Pre-processing}  %
As for the video pre-processing, we utilize Dlib \cite{King2009Dlib} to detect 68 facial landmarks (including 20 mouth landmarks), and extract the face images from video frames. We resize the face images to $256 \times 256$, and further crop each face to a fixed $160 \times 80$ size containing the lip-centered region. As for the text pre-processing, we encode the text sequence at the character level for GRID dataset and phoneme level for TCD-TIMIT dataset. And for ground truth duration extraction, we first extract the speech audio from video files, and then utilize ``Penn Phonetics Lab Forced Aligner'' (P2FA) \cite{yuan2008speaker} to get speech-to-text alignments, from which we obtain the duration of each text token for training our duration predictor in ParaLip. 

\begin{figure*}[!htb]
	\centering
	\includegraphics[width=\textwidth,trim={0cm 0.0cm 0cm 0cm}, clip=true]{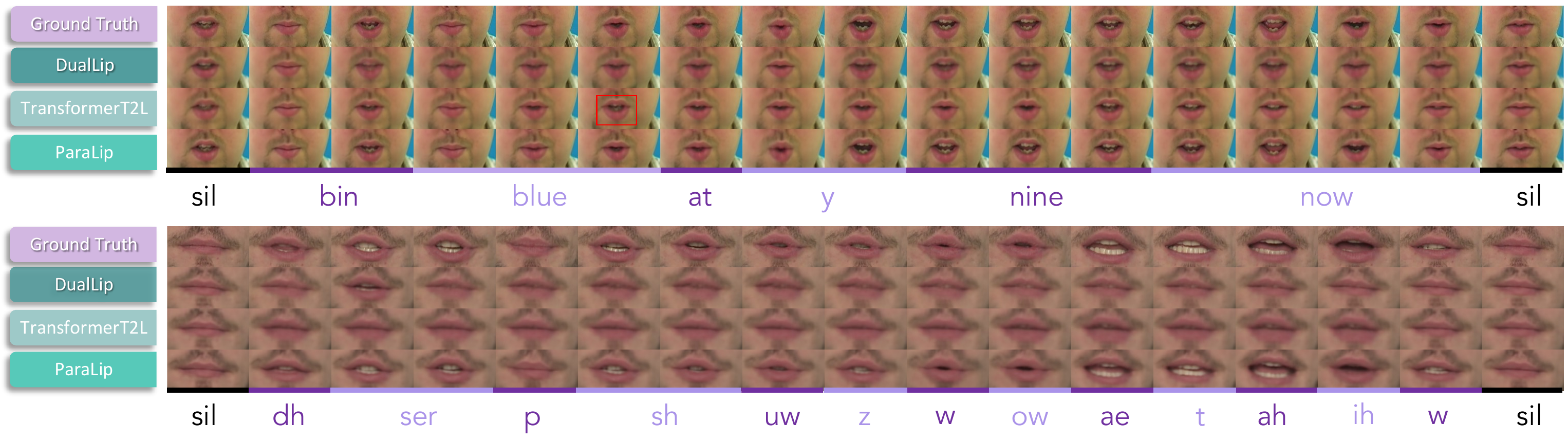}
	\small
	\caption{The qualitative comparison among AR SOTA (\textit{DualLip}), AR baseline (\textit{TransformerT2L}) and our NAR method (\textit{ParaLip}). We visualize two cases from GRID dataset and TCD-TIMIT dataset to illustrate the error propagation problem existing in AR generation and verify the robustness of \textit{ParaLip}. In the first case, the lip sequence generated from AR baseline predicts a wrong lip image (the $6$-th frame with red box), and as a result, the subsequent lip images conditioned on that image becomes out of synchronization with linguistic information and ends in chaos; \textit{DualLip} alleviates the error propagation to some degree. In the second case, both AR models perform poorly on the long-sequence dataset and generate the frames that look speechless as the time goes further.}
	\label{quali_ar_hh}
\end{figure*}

\section{Results and Analysis}   
In this section, we present extensive experimental results to evaluate the performance of \textit{ParaLip} in terms of lip movements quality and inference speedup. And then, we conduct ablation experiments to verify the significance of all proposed methods in \textit{ParaLip}.

\subsection{Quality Comparison}
We compare our model with 1) \textit{DualLip} \cite{chen2020DualLip}, which is the state-of-the-art (SOTA) autoregressive text-to-lip model based on RNN and location-sensitive attention~\cite{shen2018natural}. And 2) \textit{TransformerT2L}, an autoregressive baseline model based on Transformer~\cite{vaswani2017attention} implemented by us, which uses the same model settings with \textit{ParaLip}\footnote{Most modules and the total number of model parameters in \textit{ParaLip} and \textit{TransformerT2L} are similar.}. The quantitative results on GRID and TCD-TIMIT are listed in Table \ref{tab:grid_basic_results} and Table \ref{tab:timit_basic_results} respectively\footnote{Note that the reported results are all under the case where the ground truth (GT) duration is not provided at inference (denoted as w/o duration in \cite{chen2020DualLip}) , since there is no GT duration available in the real case.}. Note that we do not add adversarial learning on any model in Table \ref{tab:grid_basic_results} or Table \ref{tab:timit_basic_results}, since there is no adversarial learning in \textit{DualLip} \cite{chen2020DualLip}.

\begin{table}[ht]
\centering
\small
\begin{center}
\begin{tabular}{*{4}{l}}
\toprule
Methods                   & PSNR $\uparrow$       & SSIM $\uparrow$         & LMD $\downarrow$ \cr
\midrule
AR Benchmarks  & ~ & ~ & ~ \cr
\midrule
\textit{DualLip}                &  \textbf{29.13}$^\dagger$  & 0.872$^\dagger$ & 1.809 $^\dagger$ \cr
\textit{TransformerT2L}        &  26.85                      & 0.829           & 1.980            \cr
\midrule
Our Model & ~ & ~ & ~ \cr
\midrule
\textit{ParaLip}       &  28.74  & \textbf{0.875}  & \textbf{1.675} \cr

\bottomrule

\end{tabular}
\end{center}
\small
\caption{Comparison with Autoregressive Benchmarks on GRID dataset. $^\dagger$ denotes our reproduction under the case w/o GT duration at inference.}
\label{tab:grid_basic_results}

\end{table}

\begin{table}[ht]

\centering
\begin{center}
\small
\begin{tabular}{*{4}{l}}
\toprule
Methods                   & PSNR $\uparrow$       & SSIM $\uparrow$         & LMD $\downarrow$ \cr
\midrule
AR Benchmarks  & ~ & ~ & ~ \cr
\midrule 
\textit{DualLip}          &  27.38$^\dagger$  & 0.809$^\dagger$ & 2.351$^\dagger$ \cr
\textit{TransformerT2L}    &  26.89  & 0.794  & 2.763   \cr
\midrule
Our Model & ~ & ~ & ~ \cr
\midrule
\textit{ParaLip}    &  \textbf{27.64}  & \textbf{0.816} & \textbf{2.084} \cr
\bottomrule

\end{tabular}
\end{center}
\small
\caption{Comparison with Autoregressive Benchmarks on TCD-TIMIT dataset. $^\dagger$ denotes our reproduction under the case w/o GT duration at inference. }

\label{tab:timit_basic_results}
\end{table}

\subsubsection{Quantitative Comparison}
We can see that: 1) On GRID dataset (Table \ref{tab:grid_basic_results}), \textit{ParaLip} outperforms \textit{DualLip} on LMD metric, and keeps the same performance in terms of PSNR and SSIM metrics. However, on TCD-TIMIT dataset (Table \ref{tab:timit_basic_results}), \textit{ParaLip} achieves a overall performance surpassing \textit{DualLip} by a notable margin, since autoregressive models perform badly on the long-sequence dataset due to accumulated prediction error; 2) \textit{ParaLip} shows absolute superiority over AR baseline \textit{TransformerT2L} in terms of three quantitative metrics on both datasets; 3) although \textit{DualLip} outperforms AR baseline by incorporating the technique of location-sensitive attention, which could alleviate the error propagation, it is still vulnerable on long-sequence dataset.

\subsubsection{Qualitative Comparison}
We further visualize the qualitative comparison between \textit{DualLip}, \textit{TransformerT2L} and \textit{ParaLip} in Figure \ref{quali_ar_hh}. It can be seen that the quality of lip frames generated by \textit{DualLip} and \textit{TransformerT2L} become increasingly worse as the time goes further. Concretely, The lip image becomes fuzzy and out of synchronization with linguistic contents. We attribute this phenomenon to the reason that: error propagation problem is serious in AR T2L since the wrong prediction could take place at more dimensions (every pixel with three channels in generated image) and there is information loss during the down-sampling when sending the last generated lip frame to predict current one. What's worse, on TCD-TIMIT, a long-sequence dataset, \textit{DualLip} and \textit{TransformerT2L} often generate totally unsatisfying results that look like speechless video. By contrast, the lip frames generated by NAR model \textit{ParaLip} maintain high fidelity to ground truth all the while, which demonstrates the effectiveness and robustness of NAR decoding.

\subsection{Speed Comparison}
In this section, we evaluate and compare the average inference latency of \textit{DualLip}, \textit{TransformerT2L} and \textit{ParaLip} on both datasets. Furthermore, we study the relationship between inference latency and the target video length.
\subsubsection{Comparison of Average Inference Latency}
The average inference latency is the average time consumed to generate one video sample on the test set, which is measured in seconds. Table \ref{tab:latency} exhibits the inference latency of all systems. It can be found that, 1) compared with \textit{DualLip}, \textit{ParaLip} speeds up the inference by $13.09\times$ and $19.12\times$ on average on two datasets; 2) \textit{TransformerT2L} has the same structure with \textit{ParaLip}, but runs about 50\% slower than \textit{DualLip}, which indicates that it is NAR decoding manner in ParaLip speedups the inference, rather than the modification of model structure; 3) In regard to AR models, the time consumption of a single sentence increases to 0.5-1.5 seconds even on GPU, which is unacceptable for real-world application. By contrast, \textit{ParaLip} addresses the inference latency problem satisfactorily.

\begin{table}[!h]
\small
\centering
\begin{center}
\begin{tabular}{c|*{4}{c}}
\toprule
Datasets & Methods  & Latency (s)   & Speedup  \cr
\midrule
\multirow{3}*{GRID}
~ & \textit{DualLip}                & 0.299 & 1.00 $\times$ \cr
~ & \textit{TransformerT2L}                & 0.689 & 0.43 $\times$ \cr
~ & \textit{ParaLip}                         & 0.022 & 13.09 $\times$ \cr
\midrule
\multirow{3}*{TIMIT}  
~ & \textit{DualLip}                & 0.650 & 1.00 $\times$ \cr
~ & \textit{TransformerT2L}                 & 1.278 & 0.51 $\times$ \cr
~ & \textit{ParaLip}                         & 0.034 & 19.12 $\times$ \cr
\bottomrule
\end{tabular}
\end{center}
\small
\caption{The comparison of inference latency on GRID and TCD-TIMIT dataset. The computations are conducted on a server with 1 NVIDIA 2080Ti GPU.}

\label{tab:latency}
\end{table}

\begin{figure}[!h]
	\centering
	\includegraphics[width=\textwidth]{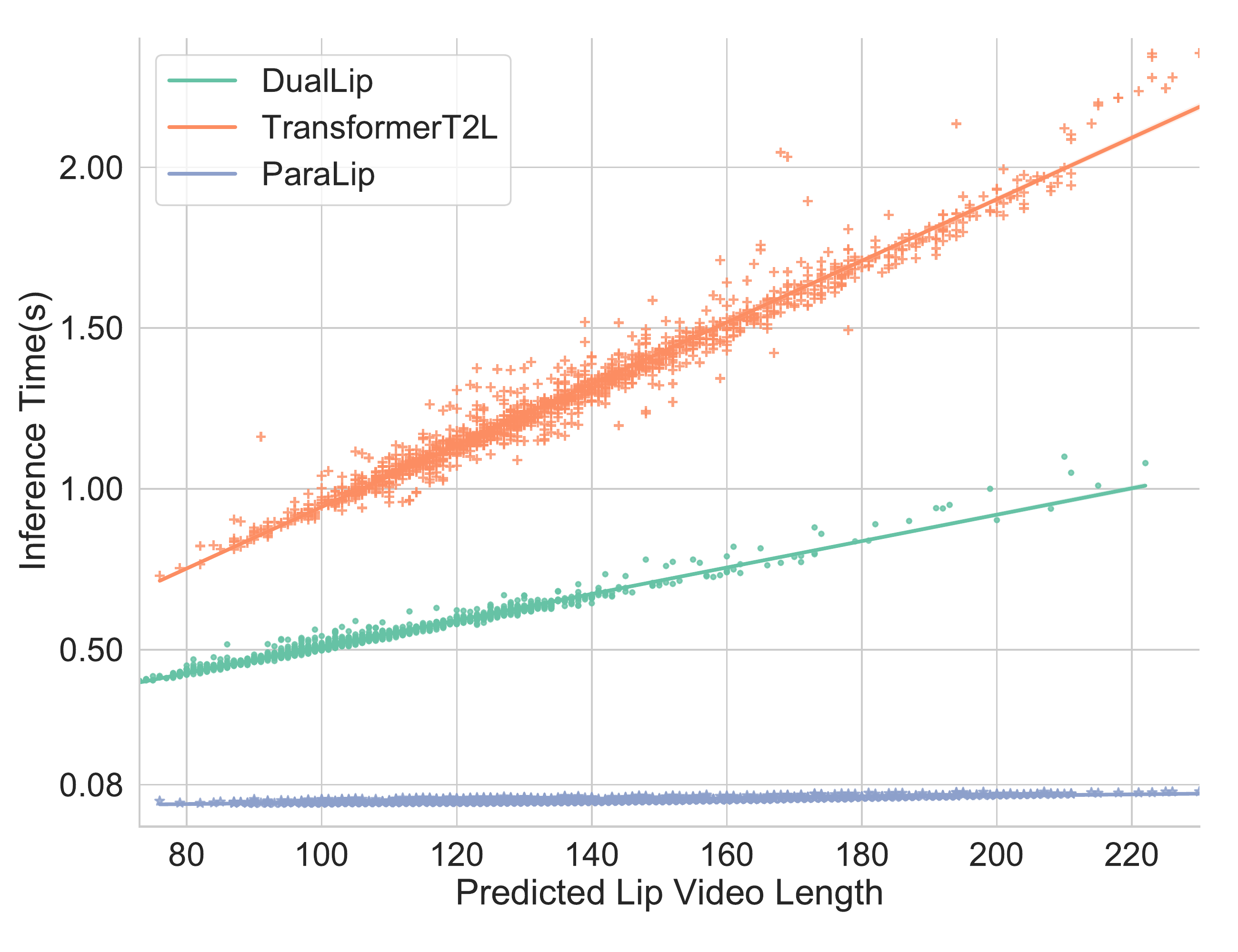}
	\small
	\caption{Relationship between inference latency (seconds) and predicted video length for \textit{DualLip}, \textit{TransformerT2L} and \textit{ParaLip}.}

	\label{seq_len_speed}
\end{figure}
\subsubsection{Relationship between Inference Latency and Video Length}
In this section, we study the speedup as the sequence length increases. The experiment is conducted on TCD-TIMIT, since its videos are not in a fixed length.
From Figure \ref{seq_len_speed}, it can be seen that 1) \textit{ParaLip} model speeds up the inference obviously due to high parallelization compared with AR models; 2) \textit{ParaLip} is insensitive to sequence length and almost holds a constant inference latency, but by contrast, the inference latency of \textit{DualLip} and \textit{TransformerT2L} increase linearly as the sequence length increases. As a result, the speedup of \textit{ParaLip} relative to \textit{DualLip} or \textit{TransformerT2L} also increases linearly as the sequence length increases. %

\subsection{Ablation Study}
\begin{figure}[!h]
	\centering
	\includegraphics[width=\textwidth]{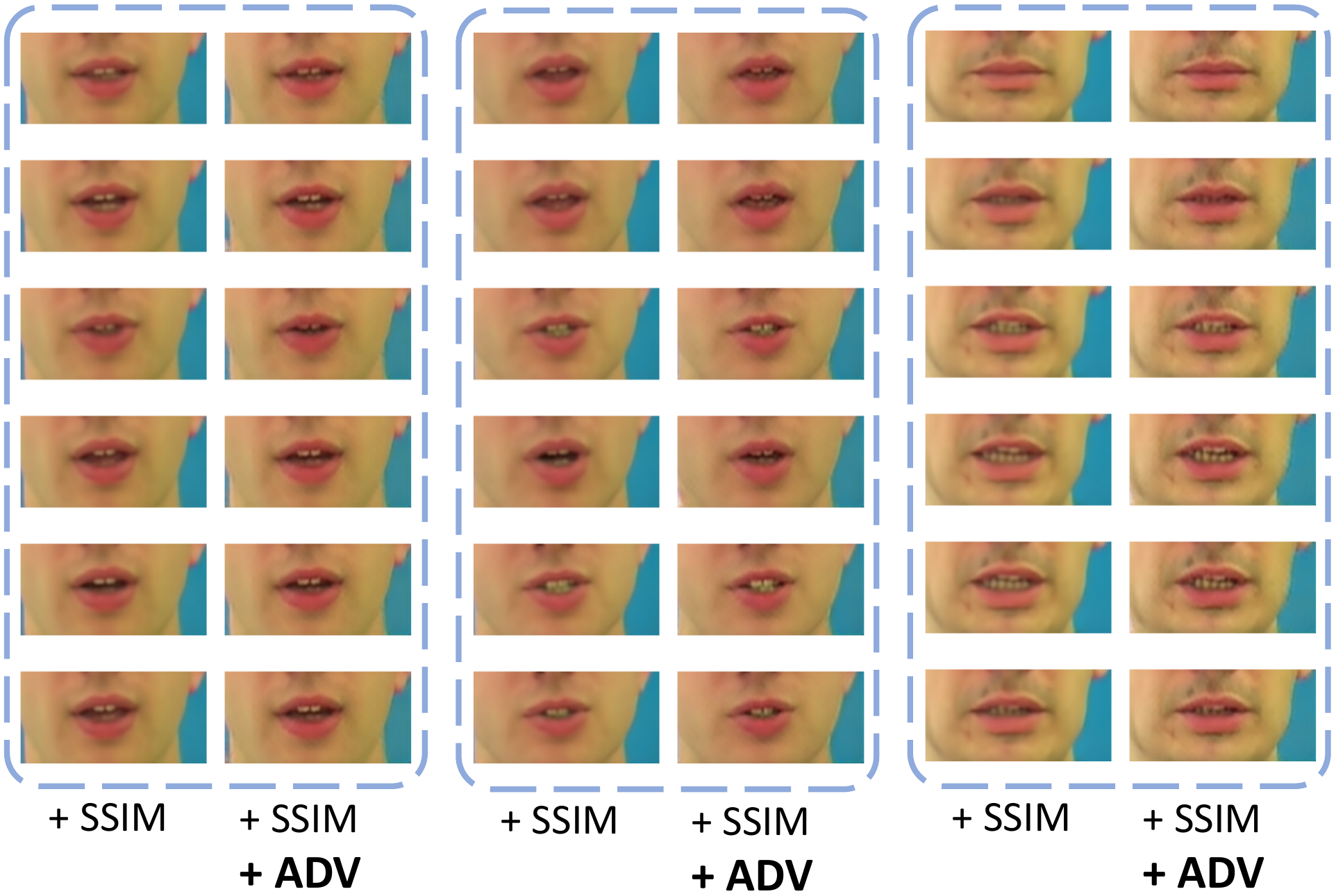}
    \small
	\caption{The qualitative evaluation for adversarial learning. ``+ADV'' tends to generate more realistic lip images.}
	\label{fig:adv_res}

\end{figure}

\begin{table}[!h]
\small
\centering
\begin{center}
\begin{tabular}{c|*{5}{c}}
\toprule
Model & PSNR $\uparrow$  & SSIM $\uparrow$ & LMD $\downarrow$ & FID $\downarrow$ \cr
\midrule
Base model       & 30.24 & 0.896 & 0.998 & 56.36\cr
+SSIM             & \underline{30.51}  &  \underline{0.906}  &  \underline{0.978} & 55.05\cr
+ADV              & 25.70  & 0.736 & 2.460 & 65.88\cr
+SSIM+ADV             & 28.36 & 0.873 & 1.077 & \underline{39.74}\cr
\bottomrule
\end{tabular}
\end{center}
\small
\caption{The ablation studies on GRID dataset. Base model is trained only with $L_1$ loss; ``+SSIM'' means adding structural similarity index loss and ``+ADV'' means adding adversarial learning to the base model. FID means Fr\'{e}chet Inception Distance metric. To focus on the frames quality, we provide the \textbf{GT} duration for eliminating the interference caused by the discrepancy of predicted length.}
\label{tab:ablation}

\end{table}
We conduct ablation experiments on GRID dataset to analyze the effectiveness of the proposed methods in our work. All the results are shown in Table \ref{tab:ablation}. Experiments show that:
\begin{itemize}
    \item Adding only SSIM loss obtains the optimal score on PSNR/SSIM/LMD (``\textit{+SSIM}'');
    \item Adding only adversarial training causes performance drop on PSNR/SSIM/LMD, which is consistent with previous works \cite{song2018talking} (``\textit{+ADV}'');
    \item Adding SSIM to model with adversarial training can greatly alleviate the detriment on PSNR/SSIM/LMD brought by adversarial training; make the GAN-based model more stable; obtain the best FID score, which means the generated lips look more realistic. (``\textit{+SSIM+ADV}'').
\end{itemize}
Previous works \cite{song2018talking} claim that 1) PSNR and SSIM cannot well reflect some visual quality; 2) adversarial learning encourages the generated face to pronounce in diverse ways, leading to diverse lip movements and LMD decrease. Although ``\textit{+SSIM+ADV}'' causes marginally PSNR/SSIM/LMD scores losses, ``\textit{+SSIM+ADV}'' obtain the best FID score and tends to generate distinct lip images with more realistic texture and local details (e.g. wrinkles, beard and teeth). The qualitative results are shown in Figure \ref{fig:adv_res}.

\section{Further Discussions}
\label{sec:further_discussion}
In the foregoing sections, we train the duration predictor using the ``GT'' duration extracted by P2FA, but it is not applicable to the case where the audio is absent even in the training set. Thus, to obtain the ``GT'' duration in this case, we tried a lipreading model with monotonic alignment searching (MAS)~\cite{tillmann1997dp,kim2020glow} to find the alignment between text and lip frames. Specifically, we 1) first trained a lipreading model by CTC loss on the training set; 2) traveled the training set and for each ($\mathcal{L}$, $\mathcal{S}$) pair, we extracted $O_{ctc} \in R^{m}$ (the CTC outputs corresponding to the label tokens) from $O_{ctc} \in R^{V}$ ($O_{ctc}$ is the original CTC outputs; V is the vocabulary size); 3) applied softmax on the extracted $O_{ctc}$ to obtain the probability matrix $P(A(s_i,l_j)) = \frac{e^{O_{ctc, lj}^{i}}}{\sum_{m} e^{O_{ctc, lj}}}$; 4) conducted MAS by dynamic programming to find the best alignment solution. This method achieves similar results with P2FA on GRID dataset, but could cause deterioration on TIMIT dataset: PSNR:27.09, SSIM:0.816, LMD:2.313. 

Theoretically, obtaining the alignment between lip frames and its transcript directly from themselves has more potential than obtaining this alignment indirectly from audio sample and its transcripts, since there are many cases when the mouth is moving but the sound hasn't come out yet (e.g. the first milliseconds of a speech sentence). This part of video frames should be aligned to some words, but its corresponding audio piece is silent, which will not be aligned to any word, causing contradictions. We think it is valuable to try more and better methods in this direction.

\section{Conclusion}
In this work, we point out and analyze the unacceptable inference latency and intractable error propagation existing in AR T2L generation, and propose a parallel decoding model ParaLip to circumvent these problems. Extensive experiments show that ParaLip generates lip movements with competitive quality compared with the state-of-the-art AR T2L model, exceeds the baseline AR model TransformerT2L by a notable margin and exhibits distinct superiority in inference speed, which provides the possibility to bring T2L generation from laboratory to industrial applications.

\section{Acknowledgments}
This work was supported in part by the National Key R\&D Program of China under Grant No.2020YFC0832505, No.62072397, Zhejiang Natural Science Foundation under Grant LR19F020006.

\bibliography{aaai22}

\appendix

\section{Implementation Detail} 
In our framework, for the TM blocks, we adopt Transformer as the basic structure. Thereinto, the model hidden size $d_{hidden}$, number of stacked layers $n_{TM}$, number of attention heads $n_{head}$ and dropout rate are set to 256, 4, 2, and 0.2 respectively.  
For the convolutional layers, the number of stacked layers and the kernel size are set to 4 and 5 by default. The window size for calculating SSIM is set to 11. As for the identity lip image, we randomly select an identity lip from target lip frames at the training stage, and use the first lip frame at inference stage.

To train ParaLip, the weights of loss functions $\lambda_1$, $\lambda_2$, $\lambda_3$ and $\lambda_4$ are set to $1$, $1$, $1$ and $5$. The batch size is set to 4 and the total number of iterations is set to $110,000$. The learning rate is set as 0.001 and 0.0001 for generator and discriminator respectively using Adam optimizer. The training is run on one RTX 2080ti GPU and our implementation is based on Pytorch Lightning. We will release our code once the paper is published.

\section{Model Size}
The model footprints of main systems for comparison in our paper are shown in Table~\ref{suptab:model_size}. It can be seen that \textit{ParaLip} has the similar learnable parameters of generator as other state-of-the-art models. 
\begin{table}[!h]
\begin{center}
    \begin{tabular}{ l | c } 
        \toprule
        \textbf{Model} &  \textbf{Param(M)} \cr
        \midrule 
        \textit{DualLip} & 41.483 \cr
        \midrule
        \textit{TransformerT2L} & 37.780 \cr
        \midrule
        \multirow{2}*{\textit{ParaLip}} 
         ~ & 35.224 (Generator)  \cr
         ~ & 14.114 (Discriminator)  \cr
        \midrule
    \end{tabular}
\end{center}
\caption{The model footprints. Param means the learnable parameters.}
\label{suptab:model_size}
\end{table}

\section{Model Details of Identity Encoder}
The details of the identity encoder (in both ParaLip and TransformerT2L) are given in Table~\ref{tab:idenc_arch}. The filter is in the format of ([h $\times$ w, channel] / stride, pad).

\begin{table}[h]
\begin{center}
\scriptsize
\begin{tabular}{ lll } 
 \toprule
 \textbf{Layer Type} & \textbf{Filters} & \textbf{Output dimensions} \\ 

 \midrule
 Conv 2D  & [$5\times5$, $16$]  $/  2, 2$  & $T\times 40 \times 80 \times 16$ \\
 \midrule                                                              
 Conv 2D  & [$5\times5$, $32$]   $/ 2, 2$  & $T\times 20 \times 40 \times 32$ \\
 \midrule                                                             
 Conv 2D  & [$5\times5$, $64$]   $/ 2, 2$  & $T\times 10 \times 20 \times 64$ \\
 \midrule  
 Conv 2D  & [$5\times5$, $128$]   $/ 2, 2$  & $T\times 5 \times 10 \times 128$ \\
 \midrule  
 Linear  & -  & $T \times 256$ \\
 \midrule  
\end{tabular}
\normalsize

\end{center}
\caption{Model details for Identity Encoder. There is a BatchNorm layer and a ReLU activation function after each Conv 2D layer.}
\label{tab:idenc_arch}
\end{table}

\section{Model Details of Discriminator in ParaLip}
The details of the discriminator in ParaLip for adversarial learning are given in Table~\ref{tab:disc_arch}. The filter is in the format of ([h $\times$ w, channel] / stride, pad).

\begin{table}[htb]
\begin{center}
\scriptsize
\begin{tabular}{ lll } 
 \toprule
 \textbf{Layer Type} & \textbf{Filters} & \textbf{Output dimensions} \\ 
 \midrule
 Conv 2D  & [$7\times7$, $32$]   $/  1, 3$  & $T\times 80 \times 160 \times 32$ \\
 \midrule
 Conv 2D  & [$5\times5$, $64$]   $/  [1,2], 2$  & $T\times 80 \times 80 \times 64$ \\
\addlinespace[0.5em]
 Conv 2D  & [$5\times5$, $64$]    $/  1, 2$  & $T\times 80 \times 80 \times 64$ \\
 \midrule                                                              
 Conv 2D  & [$5\times5$, $128$]    $/ 2, 2$  & $T\times 40 \times 40 \times 128$ \\
\addlinespace[0.5em]
 Conv 2D  & [$5\times5$, $128$]    $/ 1, 2$  & $T\times 40 \times 40 \times 128$ \\
 \midrule                                                             
 Conv 2D  & [$5\times5$, $256$]    $/ 2, 2$  & $T\times 20 \times 20 \times 256$ \\
\addlinespace[0.5em]
 Conv 2D  & [$5\times5$, $256$]    $/ 1, 2$  & $T\times 20 \times 20 \times 256$ \\
 \midrule                                                             
 Conv 2D  & [$3\times3$, $512$]    $/ 2, 1$  & $T\times 10 \times 10 \times 512$ \\
\addlinespace[0.5em]
 Conv 2D  & [$3\times3$, $512$]    $/ 1, 1$  & $T\times 10 \times 10 \times 512$ \\
 \midrule  
 Conv 2D  & [$3\times3$, $512$]    $/ 2, 1$  & $T\times 5 \times 5 \times 512$ \\
\addlinespace[0.5em]
 Conv 2D  & [$3\times3$, $512$]    $/ 1, 0$  & $T\times 3 \times 3 \times 512$ \\
 \midrule  
Conv 2D  & [$3\times3$, $512$]    $/ 1, 0$  & $T\times 1 \times 1 \times 512$ \\
\addlinespace[0.5em]
 Conv 2D  & [$1\times1$, $512$]    $/ 1, 0$  & $T\times 1 \times 1 \times 512$ \\
 \midrule  
  Conv 2D  & [$1\times1$, $1$]    $/ 1, 0$  & $T\times 1 \times 1 \times 1$ \\
 \midrule  
\end{tabular}
\normalsize

\end{center}
\caption{Model details for Discriminator. There is a LeakyReLU activation function after each Conv 2D layer.}
\label{tab:disc_arch}
\end{table}

\section{Evaluation Metrics}  
Following previous T2L generation work, we adopt PSNR, SSIM \cite{Wang2004Image} and Landmark Distance (LMD) \cite{Chen2018Lip,song2018talking} for quantitative evaluation of lip quality. PSNR and SSIM are the classical reconstruction metrics to evaluate the images quality in generated video \cite{mathieu2016deep}. LMD measures lip movement accuracy from pixel-level \cite{song2018talking}. Following \citet{chen2020DualLip}, we calculate the euclidean distance between corresponding lip landmarks of the generated lip movements and ground truth, instead of all the facial landmarks.

\end{document}